\documentclass[a4paper,journal,doublecolumn]{IEEEtran}
\usepackage{amsmath,amsfonts,amssymb}

\usepackage{graphicx, epstopdf}
\usepackage{cite}
\usepackage{url}
\usepackage[dvipsnames]{xcolor}
\usepackage[hidelinks]{hyperref} 

\usepackage{config}

\acrodef{3GPP}{3rd Generation Partnership Project}
\acrodef{AP}{Access Point}
\acrodef{API}{Application Programming Interface}
\acrodef{CCI}{Commonwealth Cyber Initiative}
\acrodef{CDF}{cumulative distribution function}
\acrodef{CWC}{Centre for Wireless Communications}
\acrodef{EVT}{Extreme Value Theory}
\acrodef{HRLLC}{Hyper-Reliable and Low-Latency Communications}
\acrodef{ITU-R}{International Telecommunication Union - Radiocommunication}
\acrodef{RIS}{Reconfigurable Intelligent Surface}
\acrodef{QoS}{Quality-of-Service}
\acrodef{SIR}{Signal to Interference Ratio}
\acrodef{SINR}{Signal to Interference and Noise Ratio}
\acrodef{SNR}{Signal to Noise Ratio}
\acrodef{UAV}{Unmanned Aerial Vehicle}
\acrodef{URLLC}{Ultra-Reliable and Low-Latency Communications}

\begin{document}

\title{Dimensioning and Optimization of Reliability Coverage in Local 6G Networks}

\author{Jacek Kibi\l{}da, Dian Echevarr\'ia P\'erez, Andr\'e Gomes, Onel L. Alcaraz L\'opez,  Arthur S. de Sena, Nurul Huda Mahmood, Hirley Alves
    \thanks{J. Kibi\l{}da is with the \ac{CCI} and the Department of Electrical and Computer Engineering, Virginia Tech, USA. Email: jkibilda@vt.edu.}
    \thanks{D. E. P\'erez, O. L\'opez,  A. S. de Sena, N. H. Mahmood, and H. Alves are with the \ac{CWC}, University of Oulu, Finland.} 
    \thanks{Andr\'e Gomes is with Rowan University, USA. Email: gomesa@rowan.edu.}
}

\maketitle


\begin{abstract} 
Enabling vertical use cases for the sixth generation (6G) wireless networks, such as automated manufacturing, immersive extended reality (XR), and self-driving fleets, will require network designs that meet reliability and latency targets in well-defined service areas. In order to establish a quantifiable design objective, we introduce the novel concept of \yale{reliability coverage}, defined as the percentage area covered by communication services operating under well-defined reliability and performance targets. Reliability coverage allows us to unify the different network design tasks occurring at different time scales, namely resource orchestration and allocation, resulting in a single framework for dimensioning and optimization in local 6G networks. The two time scales, when considered together, yield remarkably consistent results and allow us to observe how stringent reliability/latency requirements translate into the increased wireless network resource demands.
\end{abstract}

\section{Introduction} \label{sec:intro} 

Recently, the \ac{ITU-R} designated \ac{HRLLC} as one of the six driving usage scenarios for \black{IMT-2030~\cite{liu2023beginning}}. \ac{HRLLC} will be the successor to \ac{URLLC}, potentially targeting revised reliability and latency figures but also enhanced scalability, deterministic performance, and seamless integration with emerging sixth-generation (6G) wireless network technologies in support of new and upcoming vertical applications, such as automated manufacturing, immersive extended reality (XR), and self-driving fleet. However, there is the rub: despite significant research efforts and success in controlled settings~\cite{Lopez.2023}, \ac{URLLC} services have yet to translate into real-world deployments~\cite{park22_extremeURLLC}, with some studies reporting on 
 their limited attainability at a system level~\cite{maghsoudnia2024ultra}, and therefore putting into question the feasibility of \ac{HRLLC}. 

\begin{figure}[t!]
    \centering
    \includegraphics[width=1\linewidth]{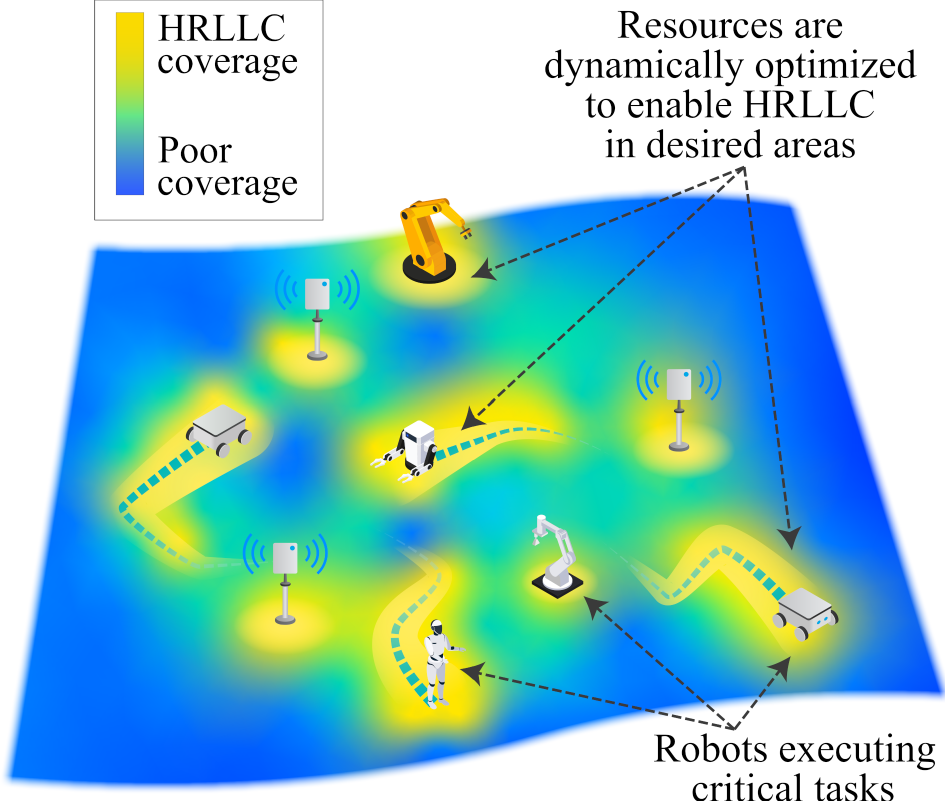}
    \caption{Visualization of the localized network coverage of reliability-based services within an automated factory scenario with autonomous vehicles and robots. 
    }
    \label{fig:visualization}
\end{figure}

In this article, we argue that one key open challenge to enable reliability and low latency-based services such as \ac{URLLC}, or more prospectively \ac{HRLLC}, lies in ensuring \black{the availability of a network that is able to meet} reliability and latency targets in \black{the} contexts specified by a vertical \black{requiring those services. Herein, we will use the term vertical to refer to a business or an organization focused on delivering highly specialized mission-critical services to a well-defined population. This, for instance, can be a} factory, a transportation \black{company}, or \black{a healthcare facility.} A common requirement \black{for a vertical}, particularly in machine-centric applications, is the need for operating across a well-defined, often local, service area~\cite{berardinelli2021extreme}. We can illustrate our argument with an example of a factory campus, as in \Fig{visualization}, where vehicles and robots operate locally and autonomously, performing tasks over delimited routes and zones that require wireless network coverage at stringent levels of reliability 
to enable seamless operations and safety. Other examples, inspired by the 6G use cases envisioned by the Next G Alliance
, can also be construed: 
i) remote surgery and health monitoring must be provided in well-defined areas of a healthcare facility or a patient location; ii) transportation networks, particularly those involving autonomous vehicles, must ensure safety and coordination across all stretches of the transport grid; and iii) emergency services must ensure timely response and coordination within a city district. 

\black{The challenge is that the widely adopted \ac{ITU-R} definition of reliability for \ac{URLLC} usage scenarios~\cite{itu2017minimum} takes a per-link reliability perspective by focusing on the success probability of transmitting a layer 2 or 3 packet between two radio communication end-points rather than that of the vertical service area. The service area perspective calls for new ways to express vertical requirements for \ac{URLLC} and overall reliability, focusing on network design targets and radio coverage.}

\black{The notion of reliability in radio coverage can be traced back to the concept of meta-distribution that defines the statistical distribution of \ac{SIR} outages over space~\cite{haenggi2021meta2}. This more granular perspective on \ac{SIR} distributions allows one to statistically describe the percentage of links that may be served at a given signal level with a pre-specified reliability, and assuming appropriately chosen statistical processes describing the network's spatial distribution. At around the same time, the notion of cell availability was introduced to quantify the percentage of the cell area where the \ac{SIR} is guaranteed to exceed a certain threshold at a pre-specified guarantee level~\cite{mendis2017achieving}. Both approaches are focused on stochastic modeling and the stochastic description of \acp{SIR} in particular. More recently, translating location-specific performance into network performance coverage was considered as a task for machine learning models that can be trained on real or synthetic datasets to produce the \emph{performance manifold}~\cite{mondal2022deep}. However, the outstanding question in those works is how to translate vertical service requirements associated with connectivity (in particular \ac{URLLC} or \ac{HRLLC}) offering to network dimensioning and optimization targets. Since \ac{SIR} (or, broadly, \ac{SINR}) is considered the \emph{first-order predictor of link reliability} and a fundamental block for any communications performance metrics~\cite{hmamouche2021new}, we will use it to build our proposed framework.}


In this article, \black{we propose the notion of \emph{reliability coverage}, which quantifies the service area (in percentage) of communication services operating under well-defined performance and link reliability targets. Subsequently, building on our prior works~\cite{gomes2023assessing,gomes2023dimensioning,perez2024evt}, we show how reliability coverage can be used to aid in network design tasks such as resource dimensioning and optimization. Here, resource dimensioning refers to the process of determining and orchestrating the necessary network resource levels (e.g., the amount of spectrum or network density), while resource optimization refers to the process of allocating and dynamically managing the available network resources (e.g., channels, rates, and power). Both processes are associated with different time-scales (i.e., the non-realtime and near-realtime control loops).} 

Extensive system-level simulations validate the effectiveness of the proposed framework in supporting \ac{URLLC}/\ac{HRLLC} in local 6G networks. Importantly, we observe remarkable consistency in results across the non-realtime and near-realtime \black{tasks}. 
First, more stringent reliability/latency and coverage targets necessitate increased spectrum supply. Second, the amount of spectrum needed will depend on the reliability \black{coverage} target and network density. In the low-reliability-coverage regime, networks of higher density require additional spectrum to meet the coverage targets. The opposite seems to hold in the high-reliability-coverage regime, whereby higher density networks offer better reliability coverage for a given spectrum amount. Similarly, when coverage is optimized, we see that additional resources, such as spectrum or transmit power, are required as the reliability requirements become more stringent. However, the accuracy of localization techniques strongly influences the performance: higher localization errors necessitate a more conservative resource \black{optimization} to account for uncertainties. 

\black{Consequently, the main contribution of this article is a unified framework for coverage dimensioning and optimization in local 6G networks to support current and future reliability and low-latency services. 
As such, the following text provides basic definitions associated with reliability coverage and an overview of our framework for reliability coverage dimensioning and optimization. It then delves into the problem of dimensioning of resource levels needed to support \ac{URLLC}/\ac{HRLLC} in a case study local network. That same case study local network is then subject to resource optimization based on reliability coverage maps. The article concludes with a discussion of open challenges and future research directions.}

\section{\black{Defining} Reliability Coverage}
 \label{sec:hrs1}
 
Before we introduce our proposed framework, let us recap the \ac{ITU-R}~definition of reliability in mobile communication services~
\cite{itu2017minimum}
: reliability \black{is} the success probability of transmitting a certain number of \black{(user-plane)} data bits within a latency deadline. 
In \ac{URLLC}, the prefix ultra refers to reliability levels often denoted as a function of \yale{nines} (where five-nines is 99.999\%) and low latency to deadlines of a few milliseconds or less. 

\black{Following from that definition, reliability can be generalized as the probability of a link associated with a user at location $x$ exceeding an arbitrary performance target $\gamma$:}
\begin{equation}
        \alpha_x(\theta, \gamma) = \Pr(\Gamma(\theta) \black{\ge} \gamma | x),
        \label{eq:link-reliability}
\end{equation}
\black{where $\Gamma(\theta)$ is the key performance metric based on \ac{SINR} affected by random phenomena (e.g., fading, blockage, and interference) and a network design configuration $\theta$ (e.g., network density, bandwidth, transmit power, number of antennas, etc.) The range of $\Gamma(\theta)$ depends generally on the application. Without loss of generality, we assume it spans non-negative real numbers, which corresponds to the conventional communications performance metrics based on \ac{SINR}, such as user-plane latency (of particular importance to \ac{URLLC} and \ac{HRLLC}), spectral efficiency, or energy efficiency~\cite{hmamouche2021new}. In more practical usage cases, such as rate selection, it can also be restricted to a set of discrete values that correspond to the discrete set of \ac{SINR} levels reported by the end-user devices.}

Consequently, we define \emph{reliability coverage} as the service area (in percentage) from which the observed reliability level is greater than or equal to a reliability requirement $\alpha^\star$. 
For a $d-$dimensional service area $\Xc\subset \Rb^d$, reliability coverage can be expressed as:
\black{\begin{equation}
\begin{split}
\eta(\theta, \gamma, \alpha^\star) &= |\Xc^\prime|/|\Xc|,
\end{split}
\label{eq:reliability-coverage}
\end{equation}
where $\Xc^\prime=\{x\in\Xc: \alpha_x(\theta, \gamma) \geq \alpha^\star\}$ is the $\alpha$-coverage set of $\Xc$, and $|\cdot|$ is the Lebesgue measure of dimension $d$. The defined $\alpha$-coverage set can be obtained directly using extensive simulations (as in \Sec{orchestration}), estimated with the help of statistical models (as in \Sec{channel-mapping}), or interpreted as the performance manifold and estimated using machine learning methods~\cite{mondal2022deep}.} 

Now, reliability coverage allows a vertical to map its requirements onto a particular network design. \black{Taking a concrete example of the }following requirements specified by the Next Generation Mobile Networks (NGMN) Alliance 
 for advanced industrial robots, \black{which} require target reliability and latency levels of $99.9999\%$ and \unit[2]{ms}. The coverage reliability requirement depends on the factory's needs and can include covering $95\%$ of the network area. 
With these requirements defined (e.g., $\alpha^\star = 99.9999\%$, $\gamma = $ \unit[2]{ms}, and $\eta^\star = 95\%$), the network can be orchestrated and resources $\theta$ allocated such that the obtained reliability coverage satisfies $\eta(\theta,\gamma,\alpha^\star) \ge \eta^\star$. 

\begin{figure*}[t]
    \centering
    \includegraphics[width=0.9\linewidth]{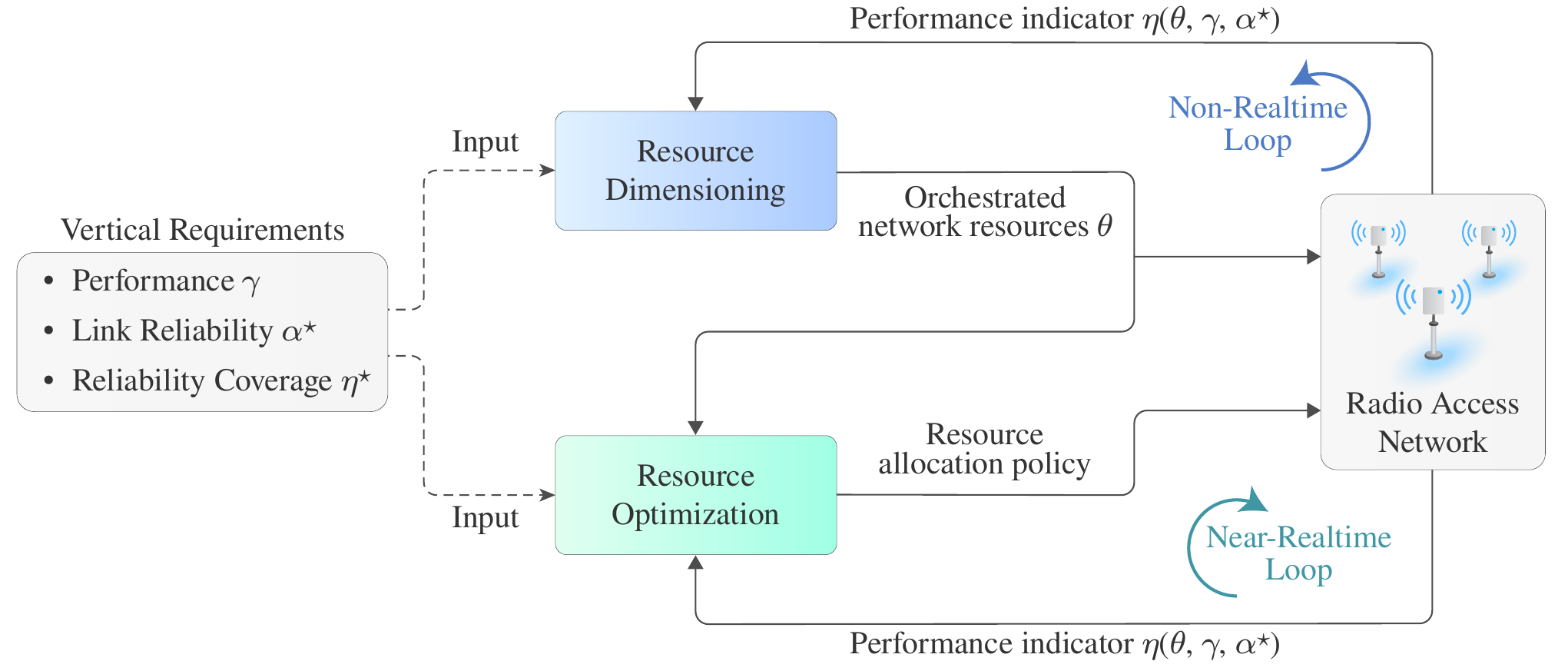}
    \caption{Framework overview: A vertical sets its targets ($\gamma,\alpha^\star,\eta^\star$) which are input to the Resource Orchestration that uses \emph{reliability coverage} to obtain the necessary network configuration, and Resource Allocation that uses these to allocate resources based on dynamic information about the channel it obtains from a \black{reliability} coverage map. Both processes result in changes monitored through control loops of varying time scales.}
    \label{fig:framework_overview}
\end{figure*}

\section{Reliability Coverage Design Framework}
 \label{sec:hrs2}
 
Our proposed framework for designing mobile networks' reliability coverage is illustrated in \Fig{framework_overview}. 
At the outset, a vertical defines its reliability service requirements in terms of its i) target performance $\gamma$, ii) target reliability level $\alpha^\star$, and iii) target reliability coverage $\eta^\star$. 

In the \emph{resource \black{dimensioning}} phase, performance targets $\gamma$ are used to \black{determine} the necessary levels of network resources $\theta$ \black{(e.g., the amount of spectrum or the density of the network).} This process involves long-scale \black{(i.e., non-real-time)} \black{estimation} to output network resource needs, and may involve long-term statistical data and models, as well as projections. The resources may come from greenfield deployments whenever a new network needs to be deployed or can be offered from the pool of resources available within an existing network in a given locality or a superposition of \black{different networks} (e.g., as in a sharing agreement between mobile operators). These can then be orchestrated into virtual network slices tailored to specific service requirements. \black{With reliability coverage, the additional challenge in the process involves the inclusion of vertical service-specific reliability and coverage targets $\alpha^\star$ and $\eta^\star$.}

In the \emph{resource \black{optimization}} phase, the determined resources are then allocated to accommodate adaptive radio resource management, resulting in a resource allocation policy. This process may involve fine-tuning and optimizing dimensioned network resources to improve the achievable rate or latency based on near-real-time data. This stage is crucial for adapting to dynamic conditions and ensuring performance requirements are met throughout the service area. \black{The process should generally take advantage of radio maps to describe location-specific characteristics of the received signal. In the case of reliability coverage, these need to be adapted to reflect the tail-end behavior of the received signal, a methodology recently explored in~\cite{kallehauge2023delivering,perez2024evt}.} Tail-end-focused radio maps provide location-specific insights into signal quality, allowing for more detailed resource allocation and the optimization of reliability coverage~\cite{kallehauge2023delivering}. As a result of both processes, different network design policies might yield long-term network statistics changes, thus calling for the triggering of non-realtime loops, followed by subsequent near-realtime optimizations.

In the following sections, we document and discuss the results of a system-level numerical study of resource \black{dimensioning} and resource \black{optimization} in application to \ac{URLLC} service provisioning over a \black{case study} local area \black{network}. Given the limited space, we only focus on the system-level aspect of reliable communications in this work, while other relevant aspects, like \black{the} distribution of failures, synchronization, etc., are left for future work.

\section{Resource \black{Dimensioning} for Reliability Coverage}\label{sec:orchestration}

In this section, we focus on \black{showcasing} the first part of the framework illustrated in \Fig{framework_overview}. 
Our goal is to devise the necessary magnitudes of network resources $\theta$ to support requirements $(\gamma, \alpha^\star, \eta^\star)$. 
Focusing on vertical scenarios of industrial networks, we consider a $200 \times 200$ meter network area and the 3GPP 3D-UMi network model in \cite{3gpp.36.873}. 
As a reference, we define our performance metric as the user-plane latency and refer to typical \ac{URLLC} requirements of transmitting 32 bytes of data within a $\gamma = $ \unit[1]{ms} deadline. 
The user-plane latency is assumed to be inversely proportional to the Shannon capacity. 
Without fine-grained information about the network deployment, at this phase, we assume a generic deployment in which \acp{AP} and user locations are modeled as a binomial point process. 
We focus on the downlink and assume that i) the transmit power is \unit[30]{dBm}, ii) \acp{AP} and mobiles are equipped with isotropic antennas installed 10 and \unit[1.5]{m} from the ground, iii) the carrier frequency is \unit[1.5]{GHz}, and iv) path loss, fading, and line-of-sight probability follow ~\cite[Tables 7.2-1 and -2]{3gpp.36.873}. 
Furthermore, we focus on interference-limited regimes, where \acp{AP} operate at the same frequency band and are constantly transmitting. 
As interference can complicate meeting \ac{URLLC} requirements, by focusing on interference-limited networks, we can estimate the upper bound of what it takes to orchestrate resources to support these services \black{(e.g., in an industrial network scenario)}. 

\begin{figure}[t]
    \centering
    \includegraphics[width=1\linewidth]{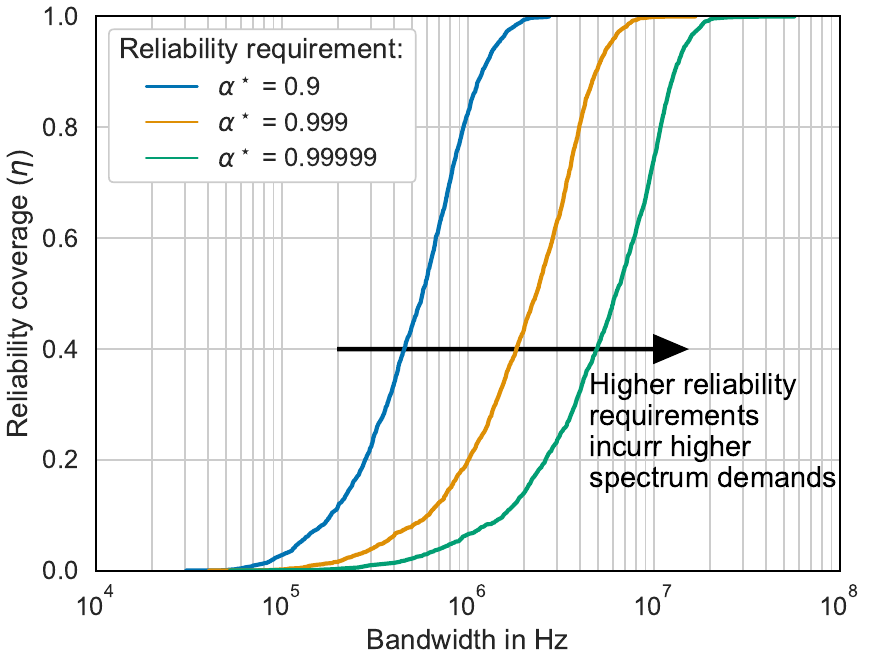}
    \caption{Reliability coverage for different $\alpha^\star$-reliability levels as a function of the bandwidth for a network density of $N = 20$ \acp{AP}. Note: $\eta \in [0, 1]$ and $\alpha^\star \in [0, 1]$.}
    \label{fig:fixed-density}
\end{figure}

For simplicity, we focus on two network resources that are at the core of any wireless systems, spectrum and network density, denoting $\theta = \{w, n\}$, where $w$ refers to the available bandwidth for communication and $n$ the number of \acp{AP}. 
First, let us consider bandwidth as a means of enabling higher reliability for a given latency. 
In our case study, additional bandwidth expands the reliability coverage. 
\Fig{fixed-density} shows the reliability coverage as a function of the bandwidth in network deployments with $n = 20$ \acp{AP} for different reliability levels. 
As we would expect, higher reliability levels $\alpha^\star$ require additional resources (bandwidth in this case) to support \black{reliability-based} services throughout the network (note how curves shift to the right); or, equivalently, that reliability coverage decreases as a function of $\alpha^\star$. 
For instance, if $w = $ \unit[10]{MHz} is available, \Fig{fixed-density} suggests that nearly the entire network service area can support reliability levels of $\alpha^\star \in \{90\%, 99.9\%\}$, but reliability coverage is limited to $\approx 80\%$ of the network service area when $\alpha^\star = 99.999\%$. 
To expand the support for $\alpha^\star = 99.999\%$ (e.g., if the required reliability coverage is $\eta^\star \ge 99\%$), \Fig{fixed-density} indicates that the service provider should consider spectrum and transmission modes that offer more than \unit[20]{MHz} of bandwidth. 

\begin{figure}[ht]
    \centering
    \includegraphics[width=1\linewidth]{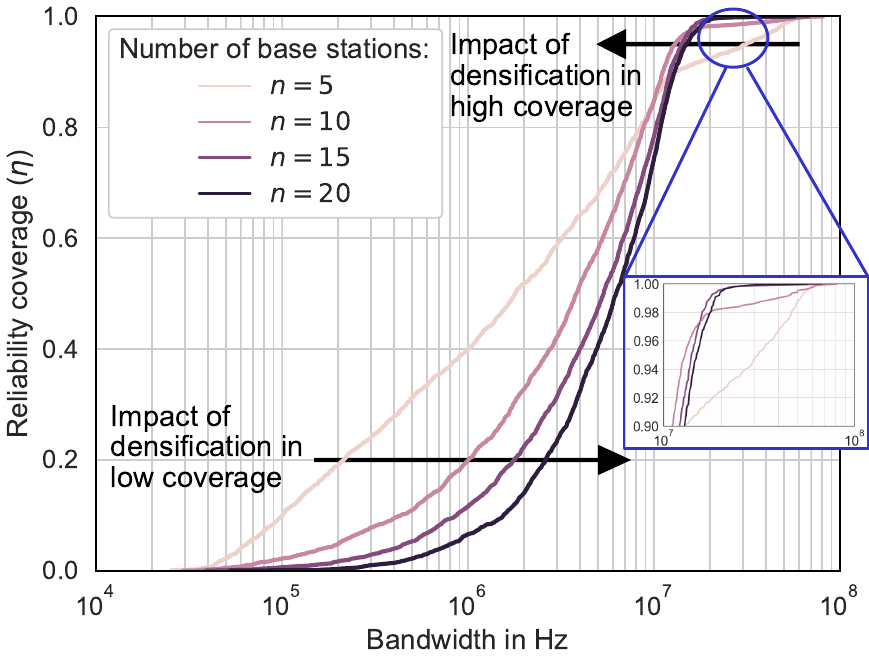}
    \caption{Reliability coverage for different network densities as a function of the bandwidth with respect to $\alpha^\star = 0.99999$ (or $99.999\%$). Note: $\eta \in [0, 1]$.}
    \label{fig:variable-density}
\end{figure}

Alternatively, the service provider can deploy more \acp{AP} to facilitate meeting \ac{URLLC} goals throughout the network. 
Intuitively, deploying more \acp{AP} can be beneficial because it reduces the pathloss between \acp{AP} and mobile users. 
In \Fig{variable-density}, we focus on the reliability target of $\alpha^\star = 99.999\%$ and consider how reliability coverage varies according to $\theta = \{w, n\}$.  
There are two important observations here. Overall, as opposed to what one might expect, deploying more \acp{AP} can increase the demand for bandwidth, implying that densification might be harmful (see how curves shift to the right). 
This is because, while densification reduces pathloss, it also increases interference in interference-limited regimes without interference control, increasing the demand for bandwidth to meet the required \ac{URLLC} performance targets. 
However, \Fig{variable-density} suggests that network density plays an important role in covering the edge and alleviates the demand for bandwidth when high reliability coverage is required. 
This is beneficial and particularly interesting in scenarios where achieving high levels of reliability coverage is desirable. 
For instance, in \Fig{variable-density}, achieving $\eta^\star = 99\%$ in sparse deployments (e.g., $n = 5$ \acp{AP}) requires tens, if not hundreds, of \unit[]{MHz} of bandwidth. 
In contrast, the same $\eta^\star = 99\%$ can be achieved with $w \approx$ \unit[20]{MHz} in deployments with $n \ge 15$ \acp{AP}. 

While the example above focuses on spectrum and density in a particular 3GPP network model, it is important to highlight that the framework does not rely on any assumption about the type of resources to be dimensioned or even the network model. 
The same methodology can be applied to dimension any set of network resources or parameters $\theta$ that can be mapped onto $\Gamma(\theta)$.

\section{Resource \black{Optimization} for Reliability Coverage} \label{sec:channel-mapping} 

The post-\black{dimensioning} phase involves dynamically managing radio resources—such as transmission channels, rates, power, and network slices—to efficiently adapt to local network conditions while meeting performance targets. 
Accurate knowledge of \ac{SINR} is critical for effective resource optimization\black{, and for the reliability coverage estimation}. This knowledge can be categorized as instantaneous \ac{SINR}, vital for real-time adjustments, and statistical \ac{SINR}, which offers insights into long-term performance and reliability under varying conditions. Both types of \ac{SINR} are complementary, with statistical \ac{SINR} being particularly important in scenarios requiring \black{ultra- and hyper-}reliability and low latency since it captures worst-case signal variations that can critically impact performance.

However, statistical \ac{SINR} can be prohibitively difficult to estimate in ultra- and hyper-reliability regimes, due to the increased data \black{collection burden~\cite{kallehauge2023delivering}}. In this framework, we propose to leverage \ac{EVT}, a powerful statistical tool for modeling the extreme tails of distributions; and radio maps, a tool to account for the spatial correlation of \ac{SINR} variations across the network. By combining \ac{EVT} for tail \ac{SINR} characterization with radio maps for spatial prediction, we can achieve a more comprehensive and practical approach to meeting the requirements of \ac{URLLC} and \ac{HRLLC} systems across local networks~\cite{perez2024evt}. Constructing accurate radio \black{reliability coverage} maps requires knowledge of multiple location-specific \ac{SINR} measurements across the coverage area. 

In order to showcase the resource \black{optimization} component of our framework, we consider \black{dimensioning}, which resulted in a deployment configuration, maintaining assumptions i) $-$ iv) from \Sec{orchestration}. Furthermore, we consider the outage requirement defined by $1-\alpha^\star$, as given in \Eq{link-reliability}. 
  We also assume precise localization of the transceivers, with five \acp{AP} deployed within the $200\times200$ meter network area and an allocated bandwidth of \unit[50]{MHz}.

\begin{figure}[h!]
    \centering
    \includegraphics[width=1\linewidth]{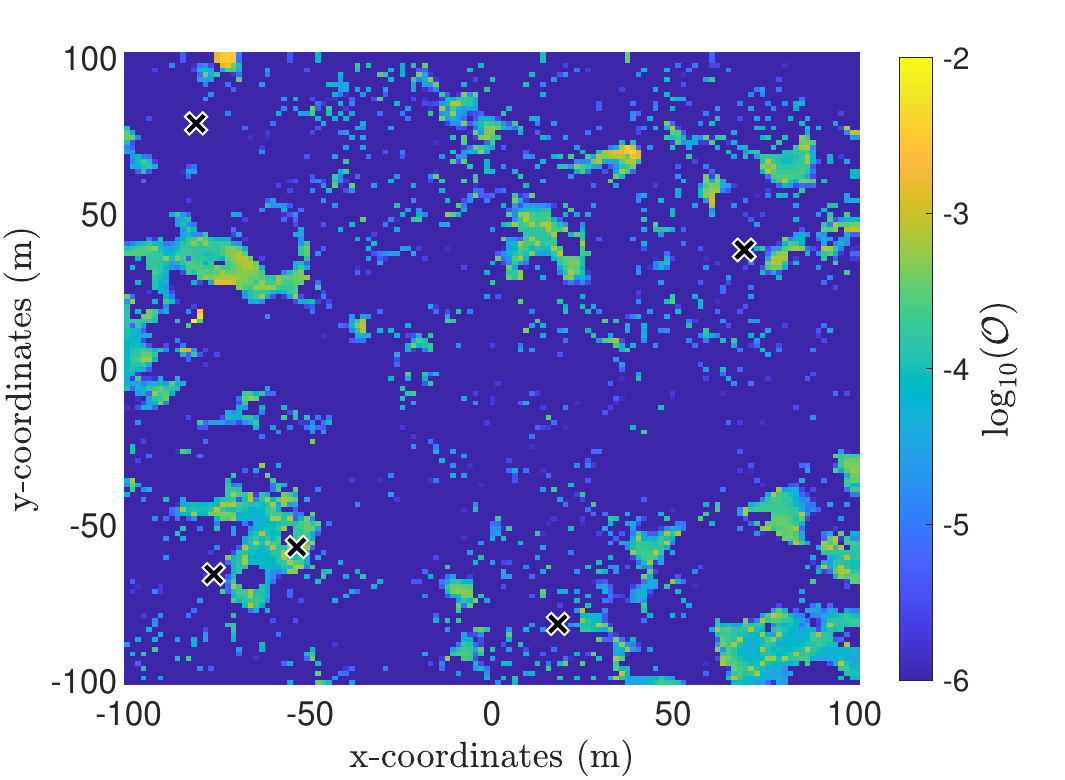}
    \caption{Map showing $\log_{10}(\mathcal{O})$, representing the achieved outage $\mathcal{O}$ with transmissions under $\gamma \leq 1 \, \mathrm{ms}$ across the coverage area. The target $10^{-3}$ is met in 99.85\% of the coverage area with a bandwidth of \unit[50]{MHz}. The markers denote the positions of \acp{AP}.
}
    \label{fig:latency_outage}
\end{figure}

\begin{figure}[h!]
    \centering
    \includegraphics[width=1\linewidth]{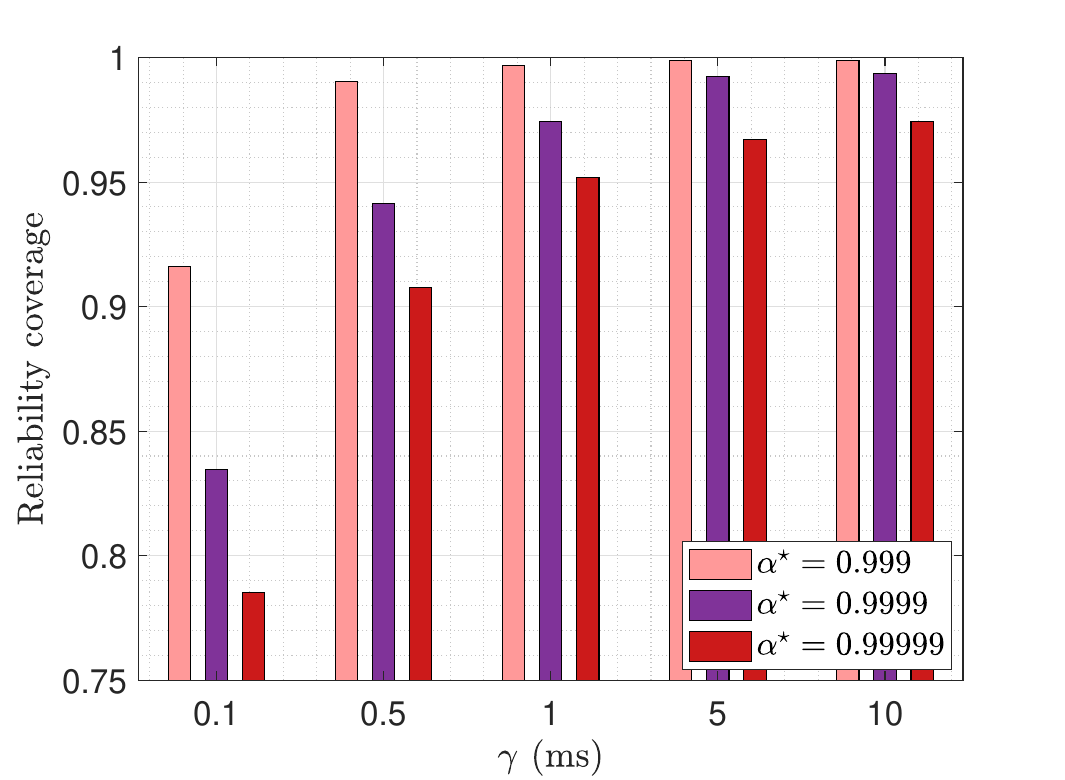}
    \caption{Reliability coverage for various latency targets $\gamma$ and $\alpha^\star$-reliability levels, assuming perfect transceiver localization and the deployment of five \acp{AP} within the coverage area. Bandwidth is fixed at \unit[50]{MHz}.
}
    \label{fig:empirical_cdf}
\end{figure}

\Fig{latency_outage} shows, on a logarithmic scale, the outage probabilities simulated with a vertical/horizontal spacing of \unit[1]{m} across the coverage area, illustrating how effectively the \black{optimized} resources meet the target $1-\alpha^\star \leq 10^{-3}$ with a \unit[1]{ms} latency constraint under the given network assumptions.
 The results indicate that the performance targets are achieved in $\eta=99.85\%$ of the service coverage. \Fig{empirical_cdf} illustrates the reliability coverage for different $\alpha$-reliability targets, latency thresholds $\gamma$, and various deployments of location-specific \ac{SINR} measurements. For less stringent reliability requirements, such as $\alpha^\star = 99.9\%$, nearly $100\%$ of the reliability coverage area meets the target outage probability with a latency below \unit[1]{ms}. However, as the reliability requirement tightens to $\alpha^\star = 99.999\%$, the reliability coverage decreases to $95.2\%$. These results align closely with those of \Fig{variable-density} for a network with five \acp{AP} and \unit[50]{MHz} of bandwidth, where the reliability coverage remains around $98\%$. Stricter latency constraints further impact reliability coverage. For instance, when $\gamma =$ \unit[0.1]{ms} with $\alpha^\star = 99.999\%$, the coverage drops significantly to $78.51\%$. Expanding the coverage area that meets these stringent requirements can be achieved, for instance, by acquiring additional spectrum or enhancing network density through additional infrastructure deployment\black{, as discussed in \Sec{orchestration}}. 

\section{Open Challenges and Future Directions} \label{sec:open-challenges} 

This article introduces the concept of \textit{reliability coverage} (i.e., the intersection of service availability with high reliability and low latency guarantees). Focusing on reliability coverage provides \black{an objective} for network design in well-defined local contexts specified by the vertical services to enable \ac{URLLC}, \ac{HRLLC}, and similar services. 
\black{In this article, two network design tasks are considered: resource dimensioning and optimization based on radio coverage maps.} Extensive simulations validate the effectiveness of the proposed framework in guaranteeing service availability with the desired reliability level. \black{The work introduced in this paper is foundational, and we would hope that it will revitalize interest in designing networks for supporting \ac{URLLC} and similar services and guide new research directions, some of which are outlined below.}

\subsection{New Tools for Reliability Coverage Analysis}
Network design tasks in the \black{ultra- and }hyper-reliability regime require large amounts of data to capture rare network events and dimension network resources accordingly. We have already shown that \ac{EVT} and Gaussian Processes can be used to alleviate the burden of data collection~\cite{kallehauge2023delivering,perez2024evt}. However, more work is needed to understand the limitations of these tools in estimating the tail-ends of the reliability coverage (i.e., when extreme reliability is expected to cover the entirety of an area). Also, advanced Monte Carlo methods for rare-event simulations, like importance sampling, subset simulation, and cross-entropy, may be leveraged to mitigate massive data collection. Integrating these methods into dimensioning and \black{optimization} frameworks poses significant challenges, including model calibration and consistency with theoretical bounds.

Meanwhile, risk management theory represents a novel direction for the design and
analysis of \ac{HRLLC} systems, including reliability coverage analysis~\cite{Lopez.2023}. More specifically, the value-at-risk and conditional-value-at-risk, which are defined as the worst loss and average worst loss, respectively, over a target horizon within a given level of confidence, can be directly exploited to model the reliability and latency constraints of \ac{HRLLC} and analyze the corresponding \black{reliability} coverage.

\subsection{Emerging Technologies for Enhanced Reliability Coverage}

\black{The resource dimensioning towards reliability coverage should take into consideration the proliferation of metasurface technologies, such as \ac{RIS}.} \ac{RIS} can help improve reliability coverage by densifying the network, engineering the wireless propagation channel, and focusing the reflected signal towards a desired location. Beyond directly improving communication performance, \acp{RIS} can be exploited to enhance the spatial characterization of wireless environments through additional optimized propagation paths. Such \ac{RIS}-enriched scattering environments can also facilitate the construction of higher-resolution
radio maps
. Conversely, radio coverage maps capturing tail-end characteristics can guide the optimal deployment of \acp{RIS} by revealing regions of low signal reliability.

The resource \black{optimization} for reliability coverage can be further enhanced through sensing, which will be an important new service for 6G networks. Environment estimation and target sensing through integrated sensing and communication techniques can result in situational awareness of the local network environment (e.g., knowledge of physical constraints, such as the presence of a blocking object) and the end users' locations. This will allow for local inference about the local channel model (and its tail-end behavior) and its posterior exploitation for better accuracy and higher resource efficiency.

In addition, collating radio sensing information with multi-modal data from other sources \black{can also help improve the estimation of reliability coverage and its optimization. Multi-modal data can take any form, from visual inputs of the network environment to building or terrain data available through open-source or commercial \acp{API}. Here, instead of \ac{EVT} and other statistical frameworks, a digital twin can aggregate this data into a latent representation of the network and its environment to perform continuous estimation of the reliability coverage and prediction of future network states. 
The digital representation can be updated with new information using continual learning approaches. At the same time, the }multi-modal data fusion \black{can} adopt split learning, a form of distributed deep learning that allows fusing multiple modalities by applying different data-specific architectures for the different modalities and then aggregating them using modality-specific weights. 

\section*{Acknowledgments}
{This work was supported by the Commonwealth Cyber Initiative (\href{https://www.cyberinitiative.org/}{cyberinitiative.org}) in Virginia, US, an investment to advance cyber R\&D, innovation, and workforce development, the National Science Foundation under grants no. 2326599 and 2421362, and the Research Council of Finland (RCF) through the projects 6G Flagship (grant no. 369116) and 6G-ConCoRSe (grant no. 359850).}

\bibliographystyle{IEEEtran}
\bibliography{ref.bib}
\vskip \xlineskip\baselineskip plus -1fil
\begin{IEEEbiographynophoto}{Jacek Kibi\l{}da} is a faculty member with \ac{CCI} and the Department of Electrical and Computer Engineering at Virginia Tech. Previously, he was a Challenge Research Fellow with Science Foundation Ireland and a Research Scientist with Wroclaw Research Center EIT+. He received his Ph.D. degree from Trinity College Dublin and his M.Sc. from Poznan University of Technology. Jacek's research focuses on modeling and design for wireless communications and networks, with an emphasis on reliability, resilience, and security.
\end{IEEEbiographynophoto}
\vskip \xlineskip\baselineskip plus -1fil
\begin{IEEEbiographynophoto}{Dian Echevarría} received the B.Sc. (2013) in Telecommunications and Electronic Engineering from the Central University of Las Villas (Cuba) and the M.Sc. (2021) in Wireless Communications Engineering from the University of Oulu (Finland). He is currently pursuing a doctoral degree at the \ac{CWC}, University of Oulu. His research focuses on radio resource management for wireless systems, emphasizing URLLC provisioning, full-duplex technology, and wireless energy transfer.
\end{IEEEbiographynophoto}
\vskip \xlineskip\baselineskip plus -1fil
\begin{IEEEbiographynophoto}{André Gomes} received his Ph.D. in Computer Engineering from Virginia Tech, US, in 2023. After his Ph.D., he was a Postdoctoral Associate with \ac{CCI}, US. Since 2024, he has been a tenure-track Assistant Professor of Computer Science at Rowan University, US. His areas of interest and research relate to wireless networking, with a focus on reliability, resilience, and controllability. He is a recipient of a best paper award at IEEE GLOBECOM 2023.  
\end{IEEEbiographynophoto}
\vskip \xlineskip\baselineskip plus -1fil
\begin{IEEEbiographynophoto}{Onel L\'opez} (S'17-M'20-SM'24) received his B.Sc., M.Sc., and D.Sc. degrees in Electrical Engineering from the Central University of Las Villas (Cuba, 2013), the Federal University of Paran\'a (Brazil, 2017), and the University of Oulu (Finland, 2020). He is a tenure-track Associate Professor at the \ac{CWC}, Oulu, Finland. He serves as an Associate Editor for several IEEE journals. His research interests include sustainable and dependable wireless communications.
\end{IEEEbiographynophoto}
\vskip \xlineskip\baselineskip plus -1fil
\begin{IEEEbiographynophoto}{Arthur S. de Sena} received his D.Sc. degree in Electrical Engineering (with distinction) from LUT University, Finland, and his M.Sc. and B.Sc. in Teleinformatics Engineering and Computer Engineering from the Federal University of Ceará, Brazil. He is currently a Postdoctoral Researcher at the University of Oulu. His research interests include signal processing, next-generation multiple access, and novel multiantenna technologies. He is an Associate Editor for IEEE Communications Letters and a recipient of the IEEE GLOBECOM 2022 Best Paper Award.
\end{IEEEbiographynophoto}
\vskip \xlineskip\baselineskip plus -1fil
\begin{IEEEbiographynophoto}{Nurul Huda Mahmood} is a senior researcher and Adjunct Professor at \ac{CWC}, University of Oulu, where he is also the coordinator for Wireless Connectivity research area and leads critical MTC research within the 6G Flagship research program. His current research focus is on resilient communications for the 6G and beyond 6G era.
\end{IEEEbiographynophoto}
\vskip \xlineskip\baselineskip plus -1fil
\begin{IEEEbiographynophoto}{Hirley Alves} 
(S’11–M’15) is an Associate Professor at \ac{CWC}, University of Oulu, where he leads Massive Wireless Automation Theme in the 6G Flagship Program. He earned his B.Sc. and M.Sc. in electrical engineering from the Federal University of Technology-Paraná, Brazil, and holds a dual D.Sc. from the University of Oulu and UTFPR. His research interests are massive and critical MTC, satellite IoT, distributed processing, and learning.
\end{IEEEbiographynophoto}

\end{document}